\documentstyle{article}
\input amssymb.sty

\renewcommand{\thefootnote}{\fnsymbol{footnote}}
\def\H{{\cal H}}
\def\id{\mathop{\rm id}}
\def\F{{\cal F}}
\def\M{\real^{p+1}}
\def\complex{{\Bbb C}}
\def\real{{\Bbb R}}
\def\inprod#1#2{\left<#1,#2\right>}
\def\pder{\partial}

\begin{document}
\begin{titlepage}
\begin{center}
\noindent 4 November 1999\hfill DIAS-STP-99-13\\

\vspace{2cm}

{\large\bf Noncommutative String Theory, the R-Matrix, and Hopf Algebras}

\vspace{1cm}

\setcounter{footnote}{0}
\renewcommand{\thefootnote}{\arabic{footnote}}

{\bf Paul Watts}\footnote{E-mail: {\tt watts@stp.dias.ie}}

\vspace{1cm}

Dublin Institute for Advanced Studies\\
School of Theoretical Physics\\
10 Burlington Road\\
Dublin 4\\
Ireland

\vspace{2cm}

{\bf Abstract}

\end{center}

\noindent Motivated by the form of the noncommutative *-product in a system
of open strings and D$p$-branes with constant nonzero Neveu-Schwarz 2-form,
we define a deformed multiplication operation on a quasitriangular Hopf
algebra in terms of its R-matrix, and comment on some of its properties.
We show that the noncommutative string theory *-product is a particular
example of this multiplication, and comment on other possible Hopf
algebraic properties which may underlie the theory.

\bigskip

\noindent PACS-95: 11.25.-w, 11.15.-q\\
\noindent MSC-91: 46L87, 57T05, 81T13

\bigskip


\end{titlepage}
\newpage
\renewcommand{\thepage}{\arabic{page}}

\section{Introduction}
\setcounter{equation}{0}

Although the subject of gauge theories in noncommutative geometry is not a
new one \cite{CL}, recently it has enjoyed something of a resurgence.  It
reappeared in the context of matrix models compactified on tori \cite{CDS},
where it was shown that such models may be reformulated as super Yang-Mills
(SYM) theories on noncommutative spacetimes when the Neveu-Schwarz (NS)
2-form $B_{ij}$ is constant and nonzero.  This is accomplished by replacing
the usual commutative multiplication of functions on the space by a
noncommutative one, denoted by *.  Subsequent studies dealt with various
aspects of noncommutative spaces \cite{HW1}-\cite{CH}.  It has also been
realised that one gets a noncommutative SYM theory for the case of a system
of open strings and D$p$-branes in a flat spacetime (with the same
condition on $B_{ij}$), provided that one not only changes all ordinary
products to *-products, but also deforms the gauge fields and their
transformations \cite{SW}.  This seems to suggest that noncommutative
geometry may be an underlying aspect of a large class of theories.

If this is so, then a mathematically consistent way of dealing with this
noncommutativity is needed, and one possibility might be to use the
language of Hopf algebras (HAs).  A HA extends the notion of an algebra by
including information about its representations (through the coproduct),
and when it is quasitriangular also gives the algebraic structure of the
modules it is represented on through its associated R-matrix, {\it i.e.}
commutation relations between module elements.  Perhaps the best-known
cases where nontrivial quasitriangular HAs play key roles are quantum
groups \cite{D,Wo}, which may be thought of consisting of matrices whose
entries are noncommuting.

It is the intent of this paper to show (in Section 3) that the *-product
mentioned above is in fact a particular case of a more general
multiplication which may be defined in terms of the R-matrix of a
particular quasitriangular HA.  This may be a clue that there is indeed a
quasitriangular HA structure to these noncommutative theories, and might
serve as a tentative first step toward finding that structure.  This could
in turn lead toward a way of formulating gauge theories on a large class of
spaces, not just commutative ones.  Some of the implications of such a
structure are commented upon in Section 4.

\bigskip

Parts of this work are somewhat pedagogical, but this is because it is
intended for an audience for whom the language of HAs may not be too
familiar; for those with some knowledge of the subject, Section 2 can be
skimmed just to determine the notation we use herein.  Others who may be
curious about HAs may find the short review useful.

\section{Hopf Algebras}
\setcounter{equation}{0}

This section is meant to be a review of both the formal aspects of what
constitutes a Hopf algebra and the explicit example where we consider the
algebra of functions over a manifold and the partial derivatives acting on
them.

\subsection{Formal Definitions}

A Hopf algebra $\H$ is an associative algebra with unit $1$ over a field
$k$ which is also equipped with a counit $\epsilon:\H\rightarrow k$, a
coproduct (or comultiplication) $\Delta:\H\rightarrow\H\otimes\H$ and an
antipode $S:\H\rightarrow\H$; the first two of these are defined to be
homomorphisms, the third an antihomomorphism, and all three satisfy the
relations
\begin{eqnarray}
\left(\Delta\otimes\id\right)\Delta(f)&=&\left(\id\otimes\Delta\right)
\Delta(f),\nonumber\\
\left(\epsilon\otimes\id\right)\Delta(f)&=&\left(\id\otimes\epsilon\right)
\Delta(f)=f,\nonumber\\
m\left(\left(S\otimes\id\right)\Delta(f)\right)&=&m\left(\left(\id\otimes
S\right)\Delta(f)\right)=\epsilon(f)1,\label{hopf}
\end{eqnarray}
where $f\in\H$, $\id$ is the identity map $f\mapsto f$ and $m$ is the
multiplication operation on $\H$ (which will usually be suppressed).  For
future reference, the first of (\ref{hopf}) is often called
`coassociativity'.

For clarity, we adopt Sweedler's notation \cite{S}, in which we write the
coproduct as $\Delta(f)=f_{(1)}\otimes f_{(2)}$, where there is an implied
summation.  For example, using this convention, the third of (\ref{hopf})
may be written as $S\left(f_{(1)}\right)f_{(2)}=f_{(1)}S\left(f_{(2)}
\right)=\epsilon(f)1$.

\bigskip

We can define the dually paired HA to $\H$, denoted $\H^*$, in the
following way: As a vector space over $k$, $\H^*$ is just the dual space to
$\H$, so there is an inner product taking $\H^*\otimes\H$ to $k$, written
as $\inprod{x}{f}$ for $x\in\H^*$ and $f\in\H$.  $\H^*$ may be given a Hopf
algebra structure by defining the multiplication, unit, coproduct, counit
and antipode on $\H^*$ via
\begin{eqnarray}
\inprod{xy}{f}&:=&\inprod{x\otimes y}{\Delta(f)},\nonumber\\
\inprod{1}{f}&:=&\epsilon(f),\nonumber\\
\inprod{\Delta(x)}{f\otimes g}&:=&\inprod{x}{fg},\nonumber\\
\epsilon(x)&:=&\inprod{x}{1},\nonumber\\
\inprod{S(x)}{f}&:=&\inprod{x}{S(f)}.\label{dual}
\end{eqnarray}
It is straightforward to show that these operations satisfy all the HA
conditions.

$\H^*$ may be thought of as an algebra of operators on $\H$ when we define
the (left) action of $x\in\H^*$ on $f\in\H$, denoted $x\cdot f$, as
\begin{equation}
x\cdot f:=f_{(1)}\inprod{x}{f_{(2)}}.
\end{equation}
Since it is easy to show that $(xy)\cdot f=x\cdot(y\cdot f)$, this is
indeed an action of elements of $\H^*$ on those of $\H$, and thus gives a
representation of $\H^*$ with $\H$ as the module.  Furthermore, there is a
sort of Leibniz rule: $x\cdot(fg)=(x_{(1)}\cdot f)(x_{(2)}\cdot g)$.  To
give $\H^*$ an interpretation, notice that $P_0(x):=x-\epsilon(x) 1$ and
$P_1(x):= \epsilon(x)1$ are projections on $\H^*$, so $\H^*=\ker
\epsilon\oplus k1$.  Note that for $x\in\ker\epsilon$, $x\cdot 1=0$, so
these may be thought of as derivatives on $\H$; elements of $k1$ just
multiply elements of $\H$ by elements of $k$.  Finally, if we have a
concept of derivatives on $\H$, we can define an integral on $\H$ as a
linear map $\int:\H\rightarrow k$ such that $\int\, x\cdot f=0$ for any
$x\in\ker\epsilon$.

\bigskip

A quasitriangular Hopf algebra is an HA $\H$ for which there is a special
invertible element $R\in\H\otimes\H$, called the R-matrix, which has the
properties
\begin{eqnarray}
\left(\tau\circ\Delta\right)(f)&=&R\,\Delta(f)\,R^{-1},\nonumber\\
\left(\Delta\otimes\id\right)(R)&=&R_{13}R_{23},\nonumber\\
\left(\id\otimes\Delta\right)(R)&=&R_{13}R_{12},\label{quasi}
\end{eqnarray}
where $\tau:f\otimes g\mapsto g\otimes f$ and the subscripts on $R$ in the
latter two above tell in which pieces of $\H^{\otimes 3}$ $R$ lives, {\it
i.e.} if $R=r_{\alpha}\otimes r^{\alpha}$ (sum implied), then $R_{13}:=
r_{\alpha}\otimes 1\otimes r^{\alpha}$.  One consequence of these
properties on the R-matrix is the fact that it must satisfy the Yang-Baxter
equation (YBE)
\begin{equation}
R_{12}R_{13}R_{23}=R_{23}R_{13}R_{12}.
\end{equation}
(Note that since $R=1\otimes 1$ satisfies all the above, all HAs are in a
sense trivially quasitriangular.)

\subsection{Hopf Algebra of Functions and Derivatives}\label{funct}

Let $\F$ be the space of functions mapping $\M$ into $\complex$.  $\F$ is
made into a commutative associative algebra in the usual way, {\it e.g.}
$(fg)(x):=f(x)g(x)$ for $f,g\in\F$ and $x\in\M$.  Furthermore, $\F$ can be
extended to an HA if we define the following on the coordinate maps $x^i$:
\begin{eqnarray}
\Delta\left(x^i\right)&:=&x^i\otimes 1+1\otimes x^i,\nonumber\\
\epsilon\left(x^i\right)&:=&0,\nonumber\\
S\left(x^i\right)&:=&-x^i.
\end{eqnarray}
Once we have the above (as well as the relations $\Delta(1)=1\otimes 1$,
$\epsilon(1)=1$ and $S(1)=1$, of course), we can use the fact that $\Delta$
and $\epsilon$ are homomorphisms and $S$ is an antihomomorphism to extend
them to all monomials of the coordinate functions and (ignoring questions
of completeness) thus to all of $\F$.

We define $\F^*$, the dually paired HA to $\F$, to be spanned by elements
$\{\pder_i|i=1,\ldots,D\}$, where the inner product between
$\pder_i\in\F^*$ and a monomial in $\F$ is
\begin{equation}
\inprod{\pder_i}{\prod_{p=1}^N\left(x^{j_p}\right)^{n_p}}=\sum_{p=1}^N
\delta_{ij_p}\delta_{n_p1}\prod_{q\neq p}\delta_{n_q0},
\end{equation}
as well as $\inprod{\pder_i}{1}=0$.  Once we have this, the HA structure of
$\F^*$ is immediate: The form of the coproduct on $\F$ tells us that $\F^*$
is commutative, the commutativity of $\F$ gives $\Delta\left(\pder_i\right)
=\pder_i\otimes 1+1\otimes\pder_i$, and $\epsilon\left(\pder_i\right)=0$
and $S\left(\pder_i\right)=-\pder_i$.

The action of $\pder_i$ on a monomial in $\F$ can be computed as well; it
will perhaps be no surprise to the reader that the result is
\begin{equation}
\pder_i\cdot\prod_{p=1}^N\left(x^{j_p}\right)^{n_p}=\sum_{p=1}^Nn_p\left(
x^{j_p}\right)^{n_p-1}\delta_{ij_p}\prod_{q\neq p}\left(x^{j_q}\right)^{
n_q}.
\end{equation}
Furthermore, for two arbitrary functions $f(x)$ and $g(x)$,
\begin{equation}
\pder_i\cdot\left(f(x)g(x)\right)=\left(\pder_i\cdot f(x)\right)g(x)+f(x)
\left(\pder_i\cdot g(x)\right),
\end{equation}
so $\F^*$ is indeed the space of partial derivatives on functions over
$\M$.  (Note: For brevity's sake, from now on we will omit the $\cdot$ when
speaking of the action of a partial derivative on a function of $x$.)

\section{The *-Product}
\setcounter{equation}{0}

Motivated by the previously-mentioned literature on noncommuting strings,
we now use the formalism just presented to introduce a new *-product on the
Hopf algebra $\H$.  We examine its properties, and then go on to show that
the noncommutative product in string theory is a particular case of this
multiplication.

\subsection{Formal Definition of the *-Product}

Suppose we have two dually paired HAs $\H$ and $\H^*$, and further suppose
that there is an R-matrix on $\H^*$.  We can thus define a new operation
$*:\H\otimes\H\rightarrow\H$ in terms of $R$ and the usual multiplication
on $\H$ via
\begin{equation}
f*g:=f_{(1)}g_{(1)}\inprod{R_{21}}{f_{(2)}\otimes g_{(2)}}.\label{star}
\end{equation}
(If $\H^*$ is trivially quasitriangular, then $f*g=fg$.)  This operation is
actually associative as a consequence of the YBE.  To see this explicitly,
we pick $f,g,h\in\H$, and first compute $(f*g)*h$:
\begin{eqnarray}
(f*g)*h&=&\left(f_{(1)}g_{(1)}\inprod{R_{21}}{f_{(2)}\otimes g_{(2)}}
\right)*h\nonumber\\
&=&f_{(1)}g_{(1)}h_{(1)}\inprod{R_{21}}{f_{(2)}g_{(2)}\otimes h_{(2)}}
\inprod{R_{21}}{f_{(3)}\otimes g_{(3)}}\nonumber\\
&=&f_{(1)}g_{(1)}h_{(1)}\inprod{R_{32}R_{31}R_{54}}{f_{(2)}\otimes g_{(2)}
\otimes h_{(2)}\otimes f_{(3)}\otimes g_{(3)}}\nonumber\\
&=&f_{(1)}g_{(1)}h_{(1)}\inprod{R_{32}R_{31}R_{21}}{f_{(2)}\otimes g_{(2)}
\otimes h_{(2)}}.
\end{eqnarray}
In going fron the first line to the second, we used the coassociativity of
$\Delta$; next, the third of (\ref{dual}), the second of (\ref{quasi}) and
a relabelling of the indices so as to stick everything into one inner
product; and the last step used the first of (\ref{dual}).  If we now
calculate $f*(g*h)$, an exactly analogous compution gives the result above
with the left argument of the inner product replaced by $R_{21}R_{31}
R_{32}$, and by using the YBE (with the first and third tensor spaces
swapped), we find the two quantities are exactly the same.  Hence,
$(f*g)*h=f*(g*h)$ and $\widehat{\H}$, the algebra constructed from (the
vector space) $\H$ and the *-multiplication is associative.  Since it turns
out that $1*f=f*1=f$, $\widehat{\H}$ is unital as well.

\bigskip

This is in general a noncommutative multiplication: Even if $\H$ itself is
commutative, $\widehat{\H}$ may not be if $R_{21}\neq R$.  To examine this
a bit more, it can be shown that the defining relations for $R$ imply that
$\left(\epsilon\otimes\id\right)(R)=\left(\id\otimes\epsilon\right)(R)=1
\otimes 1$, so that we may define a new quantity $\Theta\in\ker\epsilon\,
\otimes\ker\epsilon$ as $\Theta:=R-1\otimes 1$.  When written in terms of
$\Theta$, the *-product becomes $f*g=fg+f_{(1)}g_{(1)}\inprod{\Theta_{21}}{
f_{(2)}\otimes g_{(2)}}$.  Suppose we now define $\Theta\equiv\theta_{
\alpha}\otimes\theta^{\alpha}$ and compute $\theta^{\alpha}\cdot\left(f\,
\theta_{\alpha}\cdot g\right)$:
\begin{eqnarray}
\theta^{\alpha}\cdot\left(f\,\theta_{\alpha}\cdot g\right)&=&\theta^{
\alpha}\cdot\left(fg_{(1)}\right)\inprod{\theta_{\alpha}}{g_{(2)}}
\nonumber\\
&=&f_{(1)}g_{(1)}\inprod{\theta^{\alpha}}{f_{(2)}g_{(2)}}\inprod{
\theta_{\alpha}}{g_{(3)}}\nonumber\\
&=&f_{(1)}g_{(1)}\inprod{\Delta\left(\theta^{\alpha}\right)\otimes\theta_{
\alpha}}{f_{(2)}\otimes\Delta\left(g_{(2)}\right)},
\end{eqnarray}
where in the last step we have split up $f_{(2)}g_{(2)}$ to get the
coproduct of $\theta^{\alpha}$.  The third of (\ref{quasi}) in terms of
$\Theta$ is $\left(\id\otimes\Delta\right)(\Theta)=\Theta_{13}+\Theta_{12}
+\Theta_{13}\Theta_{12}$; if we use this, and then get rid of $\Delta\left(
g_{(2)}\right)$ by exchanging it for a multiplication in the $\H^*$
argument of the inner product, we find
\begin{equation}
\theta^{\alpha}\cdot\left(f\,\theta_{\alpha}\cdot g\right)=f_{(1)}g_{(1)}
\inprod{\Theta_{21}+(1\otimes\kappa)R_{21}}{f_{(2)}\otimes g_{(2)}},
\end{equation}
where $\kappa:=\theta^{\alpha}\theta_{\alpha}$.  Let us now suppose that
there is a kind of `tracelessness' condition on $\Theta$, and $\kappa$
vanishes.  We must stress that this is purely an assumption, but if it does
in fact hold, then we conclude that
\begin{equation}
f*g=fg+\theta^{\alpha}\cdot\left(f\,\theta_{\alpha}\cdot g\right),
\end{equation}
in other words, $f*g$ and $fg$ differ by a `total derivative', since
$\theta^{\alpha}\in\ker\epsilon$.  It follows that if there is an integral
$\int$ defined over $\H$, $\int\,f*g=\int\,fg$.

\bigskip

A comment on the above: The above can be easily extended to the case where
we have a $N\times N$ matrix-valued Hopf algebra, {\it i.e.} $\H\otimes
M_N(k)$, in which case the *-product includes matrix multiplication as
well: $(f*g)^i{}_j=f^i{}_k*g^k{}_j$.  Note that this is {\em not} the same
as a quantum group, since $\H$ is not being thought of as the set of
functions over the group manifold.  Thus, when we take coproducts {\it et
al.}, the `matrix part' is unaffected.

Also, the *-product is not unique: If $R_{21}$ is replaced by $R^{-1}$, the
new * is still associative, and the other results also follow with suitable
modifications.  In the case where the HA is {\em triangular}, $R_{21}=R^{
-1}$ by definition, and the two different *-products coincide.

\subsection{Noncommutative String Theory}

Recently \cite{SW}, it has been shown that when one considers open strings
and D$p$-branes in a space with metric $g_{ij}$ and constant nonvanishing
Neveu-Schwarz 2-form $B_{ij}$ ($i,j=0,\ldots,p$), the theory can be
reformulated as a SYM theory on a space where the coordinates no longer
commute, but instead satisfy the deformed relation
\begin{equation}
x^i*x^j-x^j*x^i=i\theta^{ij},\label{x-comm}
\end{equation}
where
\begin{equation}
\theta^{ij}:=-\left(2\pi\alpha'\right)^2\left(\frac{1}{g+2\pi\alpha'B}B
\frac{1}{g-2\pi\alpha'B}\right)^{ij}.
\end{equation}
More generally, it is shown that if $f(x)$ and $g(x)$ are (matrix-valued)
functions, the noncommutative product between two functions $f(x)$ and
$g(x)$ is
\begin{equation}
f(x)*g(x):=\left.e^{\frac{i}{2}\theta^{ij}\frac{\pder}{\pder\xi^i}\frac{
\pder}{\pder\zeta^j}}f(x+\xi)g(x+\zeta)\right|_{\xi=\zeta=0}.\label{fg-comm}
\end{equation}
The action then can be expressed in SYM form provided that all ordinary
multiplications are replaced by this *-product, the `closed string metric'
$g_{ij}$ is replaced by the `open string metric' $G_{ij}:=g_{ij}-\left(
2\pi\alpha'\right)^2\left(Bg^{-1}B\right)_{ij}$ and the gauge field $A_i$
is replaced by $\widehat{A}_i$, which depends on both $A_i$ (and its
derivatives) and $\theta^{ij}$.

If we compare (\ref{fg-comm}) to (\ref{star}), it suggests that a candidate
for the R-matrix is
\begin{equation}
R=e^{-\frac{i}{2}\theta^{ij}\pder_i\otimes\pder_j},\label{R-matrix}
\end{equation}
but we must make sure it satisfies all the appropriate relations.  Since
$\F^*$ is commutative, and $\Delta\left(\pder_i\right)$ is symmetric, it is
evident that the first of (\ref{quasi}) is satisfied.  To check the second
of (\ref{quasi}), note that
\begin{equation}
\Delta\left(\prod_{\ell=1}^k\pder_{i_{\ell}}\right)=\sum_{\sigma}\sum_{
\ell\leq k}\frac{1}{\ell!(k-\ell)!}\pder_{i_{\sigma(1)}}\ldots\pder_{i_{
\sigma(\ell)}}\otimes\pder_{i_{\sigma(\ell+1)}}\ldots\pder_{i_{\sigma(k)}},
\end{equation}
where $\sigma$ is a permutation of $(1,\ldots,k)$.  Therefore,
\begin{eqnarray}
\left(\Delta\otimes\id\right)(R)&=&\sum_{k=0}^{\infty}\frac{(-i)^k}{2^kk!}
\left(\prod_{p\leq k}\theta^{i_pj_p}\right)\Delta\left(\prod_{\ell\leq k}
\pder_{i_{\ell}}\right)\otimes\left(\prod_{q\leq k}\pder_{j_q}\right)
\nonumber\\
&=&\sum_{k=0}^{\infty}\sum_{\ell\leq k}\frac{(-i)^k}{2^kk!\ell!(k-\ell)!}
\left(\prod_{p\leq k}\theta^{i_pj_p}\right)\sum_{\sigma}\left(\prod_{m=1}^{
\ell}\pder_{i_{\sigma(m)}}\right.\nonumber\\
&&\left.\otimes\prod_{n=\ell+1}^k\pder_{i_{\sigma(n)}}\right)\otimes
\left(\prod_{q\leq k}\pder_{j_q}\right)\nonumber\\
&=&\sum_{\ell=0}^{\infty}\frac{(-i)^{\ell}}{2^{\ell}\ell!}\left(\prod_{p
\leq\ell}\theta^{i_pj_p}\pder_{i_p}\right)\otimes\left[\sum_{k\geq\ell}
\frac{(-i)^{k-\ell}}{2^{k-\ell}(k-\ell)!}\right.\nonumber\\
&&\left.\left(\prod_{n=\ell+1}^k\theta^{i_nj_n}\pder_{i_n}\right)\otimes
\left(\prod_{q\leq k}\pder_{j_q}\right)\right],
\end{eqnarray}
where in the last step we have switched the sums over $k$ and $\ell$, and
also used the commutativity of the partial derivatives in the third tensor
space to get rid of $\sigma$ (picking up a $k!$ in the process).  If we now
split the product in the third space into two, one from 1 to $\ell$ and the
other from $\ell+1$ to $k$, and let $k-\ell=r$, then we get
\begin{eqnarray}
\left(\Delta\otimes\id\right)(R)&=&\sum_{\ell,r=0}^{\infty}\frac{(-i)^{
\ell+r}}{2^{\ell+r}\ell!r!}\left(\prod_{p\leq\ell}\theta^{i_pj_p}\pder_{
i_p}\right)\otimes\left(\prod_{n\leq r}\theta^{i'_nj'_n}\pder_{i'_n}
\right)\nonumber\\
&&\otimes\left(\prod_{q\leq\ell}\pder_{j_q}\prod_{q'\leq r}\pder_{j'_{q'}}
\right)\nonumber\\
&=&\left[\sum_{\ell=0}^{\infty}\frac{(-i)^{\ell}}{2^{\ell}\ell!}\left(
\theta^{ij}\pder_i\otimes 1\otimes\pder_j\right)^{\ell}\right]\left[\sum_{
r=0}^{\infty}\frac{(-i)^r}{2^rr!}\left(1\otimes\theta^{i'j'}\pder_{i'}
\otimes\pder_{j'}\right)^r\right]\nonumber\\
&=&R_{13}R_{23}.
\end{eqnarray}
A very similar calculation confirms that the last of (\ref{quasi}) holds as
well, so (\ref{R-matrix}) is in fact an R-matrix, and $\F$ a
quasitriangular HA (in fact, since $R_{21}=R^{-1}$, it is triangular).  The
YBE is therefore satisfied by this $R$, and thus * is associative, as
proven above.

To check that this R-matrix gives us the correct commutation relation
(\ref{x-comm}), we simply compute $x^i*x^j$:
\begin{eqnarray}
x^i*x^j&=&\left(x^i\right)_{(1)}\left(x^j\right)_{(1)}\inprod{R_{21}}{
\left(x^i\right)_{(2)}\otimes\left(x^j\right)_{(2)}}\nonumber\\
&=&x^ix^j\inprod{R_{21}}{1\otimes 1}+x^i\inprod{R_{21}}{1\otimes x^j}+x^j
\inprod{R_{21}}{x^i\otimes 1}\nonumber\\
&&+\inprod{R_{21}}{x^i\otimes x^j}\nonumber\\
&=&x^ix^j+\frac{i}{2}\theta^{ij}.
\end{eqnarray}
Therefore, by switching $i$ and $j$ and subtracting, we recover
(\ref{x-comm}).  (\ref{fg-comm}) holds almost by definition, since, from
(\ref{star}), $f*g$ is simply the product between the action of the first
tensor space of $R_{21}$ on $f$ and the action of the second on $g$.

This $R$ also satisfies the `tracelessness' condition: If we subtract
$1\otimes 1$ from $R_{21}$ and multiply the two tensor product spaces
together, we immediately get $\kappa$, which is evidently zero due to the
asymmetry (and constancy) of $\theta^{ij}$ and commutativity of the partial
derivatives, so $f*g$ and $fg$ differ by a total derivative.  And since the
original multiplication was commutative to begin with, $\int{\rm tr}\,(f*g)
=\int{\rm tr}\,(g*f)$.

\section{Noncommutative Gauge Theories}
\setcounter{equation}{0}

We now make some comments on noncommutative gauge theories and how they may
or may not relate to HAs.

\subsection{Algebraic Structure}

In Section \ref{funct}, we showed that for the commutative case, there is a
HA structure to both functions and derivatives on $\M$.  However, even
though the noncommutative $\widehat{\F}$ is an associative unital algebra,
it is {\em not} a HA, as can be seen from the following: As we proved, the
unit in $\F$ is the unit in $\widehat{\F}$, so that if $\widehat{\F}$ is a
HA and has a counit $\widehat{\epsilon}$, then the fact that this counit is
a homomorphism from $\widehat{\F}$ to $\complex$ means
$\widehat{\epsilon}(1)=1$.  Now, consider $x^i*x^j-x^j*x^i$; the counit
will map this to zero.  But this commutator is $i\theta^{ij}1$, so we have
a contradiction.  Therefore, $\widehat{\F}$ cannot be a HA.

This is no real surprise.  In the first place, the requirements necessary
for a space to be a HA are very restrictive, and in general there is no
reason to expect an arbitrary algebra to also be a HA.  Furthermore,
although $\widehat{\F}^*$ has a coalgebra structure (a counit and a
coproduct) due to the fact that $\widehat{\F}$ is unital and associative,
there is no `deformed derivative'.  We can certainly define an inner
product between the two spaces, but due to the lack of a coproduct on
$\widehat{\F}$, there is no action of $\widehat{\F}^*$ on it, and thus no
concept of derivative.  This is borne out by the fact that we must use the
ordinary partial derivative to define the noncommutative SYM field
strength, via $\widehat{F}_{ij}=\pder_{[i}\widehat{A}_{j]}-i\widehat{A}_{
[i}*\widehat{A}_{j]}$ \cite{SW}; if a deformed derivative were available,
it would be the more natural choice, but this is not the case.

However, although there is no interpretation of elements of $\widehat{\F}^
*$ as objects with a {\em local} action on $\widehat{\F}$, it might still
be possible to interpret them as {\em nonlocal} operators.  As an
illustration of this, consider the case of the 2-dimensional quantum
hyperplane: The coordinates $x,y$ generate an algebra (the functions on the
plane) modulo the commutation relation $xy=qyx$, and the `derivatives'
$\pder_x,\pder_y$ act on a function $f(x,y)$ (ordered so that all $x$s
appear to the left of all $y$s) as
\begin{eqnarray}
\pder_xf(x,y)&=&\frac{f(q^{-2}x,y)-f(x,y)}{\left(q^{-2}-1\right)x},
\nonumber\\
\pder_yf(x,y)&=&\frac{f(q^{-2}x,q^{-2}y)-f(q^{-2}x,y)}{\left(q^{-2}-1
\right)y},
\end{eqnarray}
where $q\in\real$ \cite{WZ}.  Note that as $q\rightarrow 1$, these become
ordinary derivatives, but otherwise they are nonlocal difference operators.
The string case could be similar, with $\theta^{ij}$ playing the role of
$\ln q$.  This granularisation of the spacetime might explain the absence
of small instanton singularities \cite{NS,SW} in the noncommuting theory,
by smearing out such objects over more than one point.

\subsection{Gauge Fields and Hopf Algebras}

The fact that the *-product can be defined in terms of an abstract HA and
includes the noncommutative string theory case hints at the possibility of
describing the entire theory using a quasitriangular HA, where the R-matrix
depends on $\theta^{ij}$.  However, this is certainly not sufficient, since
we have not considered the gauge field $\widehat{A}_i$.  We have also not
addressed the matter of the map $A_i\mapsto\widehat{A}_i$ which allow us to
cast the action in SYM form.  We now address both of these issues.

$\theta^{ij}$ is inherently a HA parameter; it appears in the R-matrix and
therefore describes the HA structure of $\F$ and $\F^*$.  This can be seen
either explicitly, as in the relation $(\tau\circ\Delta)(x)=R\Delta(x)R^{
-1}$ on any $\H^*$, or via the commutation relations of elements of $\H$,
which may be expressed as
\begin{equation}
gf=\inprod{R}{f_{(1)}\otimes g_{(1)}}\,f_{(2)}g_{(2)}\,\inprod{R^{-1}}{
f_{(3)}\otimes g_{(3)}}.
\end{equation}
So, motivated by these facts, it therefore seems reasonable to conjecture
that {\em all} $\theta$-dependence in the theory is in the HA structure of
$\F$ and $\F^*$.

If this is true, then the $\theta$-dependence in the noncommuting theory
must arise from the underlying HA describing the commutative theory.  This
gives us a bit of information about the gauge fields: We know that the
change of variables between the gauge fields of the two theories involves
$\theta$ \cite{SW}, which means some sort of HA-derived operation is
involved in going from one to the other.  Since, as we proved in the
previous section, the noncommutative algebra $\widehat{\F}$ is not a HA,
there must be some element $W\in\F$ with given HA properties (coproduct,
{\it etc.}) which is related to both $A_i$ and $\widehat{A}_i$.  The
assumption that all $\theta$-dependence is in the HA structure and not in
$\F$ (as a vector space) itself leads to the conclusion that $W$ is
independent of $\theta$, and thus must be related to $A_i$, since this is
the gauge field for the $\theta^{ij}=0$ case.  If we also assume that the
only dependence on the open string metric $G_{ij}$ is from constructing
Lorentz-invariant quantities in the integral, {\it e.g.} $\int {\rm
d}^{p+1}x\,\sqrt{G}G^{ij}\alpha_i\beta_j$, then $W$ should also be
$G$-independent.

At this writing, we do not know precisely what this element might be, but a
natural candidate would be the Wilson line (which explains why we call it
$W$): Recall that the Wilson line $W_{C\left(x_0,x\right)}$ is given by
\begin{equation}
W_{C\left(x_0,x\right)}:={\rm P}e^{i\int_{C\left(x_0,x\right)}A},
\end{equation}
where $C\left(x_0,x\right)$ is a path going from $x_0$ to $x$.  It depends
on the gauge field of the commutative theory, and is independent of
$\theta$ and $G$, so it fits the criteria we just outlined.  We might also
be able to obtain $\widehat{A}_i$ in the following way: Our proposed $W$
must relate both gauge fields in a HA-dependent way, but be independent of
$\theta$.  The relation of $W_C$ to $A_i$ is through the multiplication on
$\F$, via the path-ordered exponent.  We could therefore {\em define}
$\widehat{A}_i$ to be that function for which the same $W_C$ can be written
as a path-ordered exponential as well, but this time using the *-product
instead (denoted $\widehat{e}$).  In other words,
\begin{eqnarray}
W_{C\left(x_0,x\right)}&=&{\rm P}\widehat{e}^{\,i\int_{C\left(x_0,x\right)}
\widehat{A}}\nonumber\\
&=&1+i\int_{C\left(x_0,x\right)}\widehat{A}-\int_{C\left(x_0,x\right)}
\left(\widehat{A}*\int_{C\left(x_0,x_1\right)}\widehat{A}\right)+\ldots
\end{eqnarray}
Then the condition
\begin{equation}
\frac{\pder}{\pder\theta^{ij}}{\rm P}\widehat{e}^{\,i\int_{C\left(x_0,x
\right)}\widehat{A}}=0
\end{equation}
would give a differential equation involving $\widehat{A}_i$, $\theta^{ij}$
and $W_C$, which we could presumably solve to find $\widehat{A}_i$ as a
function of $\theta^{ij}$ and (derivatives of) $A_i$.  Our first
calculations show enough similarities to the first of Equation (3.5) of
\cite{SW} to be encouraging.

As for the gauge transformation, the same idea applies: A finite
transformation on $W_C$ given by a unitary matrix $U(x)=e^{i\lambda(x)}$
gives $U(x)W_{C\left(x_0,x\right)}U^{-1}\left(x_0\right)$.  This should be
the same as if we started with $W_C$ in terms of * and $\widehat{A}_i$, and
then transformed by $U$ expressed in terms of * and a new matrix
$\widehat{\lambda}$ via $U=\widehat{e}^{\,i\widehat{\lambda}}$.  Then by
solving
\begin{equation}
\frac{\pder}{\pder\theta^{ij}}\left[\widehat{e}^{\,i\widehat{\lambda}(x)}*
\widehat{e}^{\,i\int_{C\left(x_0,x\right)}\widehat{A}}*\widehat{e}^{\,-i
\widehat{\lambda}\left( x_0\right)}\right]=0,
\end{equation}
we would obtain the second of (3.5) in \cite{SW}.

So the sought-after $W\in\F$ could be related to the Wilson loop $W_C$,
even though we do not claim to have offered anything more than a vague
justification for this feeling.  We have not said anything about the HA
properties of $W$ (although the HA enters in going from $e$ to
$\widehat{e}$, via *), nor have we said how the curve $C\left(x_0,x\right)$
would be chosen, since we have taken it to be completely arbitrary.  And
perhaps most importantly, we have not considered what the action might be
as a function of $W$ (and $R$) such that we ultimately end up with a SYM
form when we go to the noncommuting spacetime.  Regardless, there is enough
evidence to consider our Wilson line guess as reasonable.

\bigskip

Everything we have done above has been for the specific case of a
noncommutative string, where we started with a commutative space ($\M$);
formulating it in the HA language as we propose would also be a way of
coming up with gauge theories where the original space might be
noncommutative.  Steps in this direction have been made when the gauge
group (and possibly also the spacetime) is deformed \cite{W}, and this may
mean there is some hope of success for the present problem.

\section{Conclusions}

We have shown that the product appearing in noncommutative string theory is
simply a specific case of one which may be defined in terms of a
quasitriangular HA.  This fact has lead us to speculate that it may be
possible to relate an arbitrary noncommutative gauge theory to a
quasitriangular HA in this way, and we have commented on some ways in which
this may be done.  We hope to address this possibility in future work.

\section*{Acknowledgements}

I would like to thank Lochlainn \'O Raifeartaigh, Jan Pawlowski and J\"org
Teschner for their helpful comments, suggestions and advice.

\end{document}